\documentclass[showpacs,endfloats,superscriptaddress,twocolumn]{revtex4}

\usepackage[latin1]{inputenc}
\usepackage{graphicx}%
\usepackage{amsmath}

\makeatletter
\def\btt#1{\texttt{\@backslashchar#1}}%
\DeclareRobustCommand\bblash{\btt{\@backslashchar}}%
\makeatother

\newcommand{\bomega}{\mbox{\boldmath $\boldmath{\omega}$\unboldmath }}

\begin{document}


\title{Dynamics of simple liquids at heterogeneous surfaces : Molecular Dynamics
simulations and hydrodynamic description }
\author{C\'ecile Cottin-Bizonne}
\affiliation{School of Chemical and Biomolecular Engineering,
Cornell University, Ithaca, NY 14853} \affiliation{Laboratoire de
Physique de la Mati\`ere Condens\'ee et Nanostructures, CNRS and
Universit\'e Lyon I,  43 Boulevard du 11 Novembre, 69622
Villeurbanne Cedex}
\author{Catherine Barentin}
\affiliation{Laboratoire de Physique de la Mati\`ere Condens\'ee
et Nanostructures, CNRS and  Universit\'e Lyon I,  43 Boulevard du
11 Novembre, 69622 Villeurbanne Cedex}
\author{\'Elisabeth Charlaix}
\affiliation{Laboratoire de Physique de la Mati\`ere Condens\'ee
et Nanostructures, CNRS and  Universit\'e Lyon I,  43 Boulevard du
11 Novembre, 69622 Villeurbanne Cedex}
\author{Lyd\'eric Bocquet}
\email{lbocquet@lpmcn.univ-lyon1.fr} \affiliation{Laboratoire de
Physique de la Mati\`ere Condens\'ee et Nanostructures, CNRS and
Universit\'e Lyon I,  43 Boulevard du 11 Novembre, 69622
Villeurbanne Cedex}
\author{Jean-Louis Barrat}
\affiliation{Laboratoire de Physique de la Mati\`ere Condens\'ee
et Nanostructures, CNRS and  Universit\'e Lyon I,  43 Boulevard du
11 Novembre, 69622 Villeurbanne Cedex}
\date{\today}

\begin{abstract}
In this paper we consider the effect of surface heterogeneity on
the slippage of fluid, using two complementary approaches. First,
MD simulations of a corrugated hydrophobic surface have been
performed. A dewetting transition, leading to a super-hydrophobic
state, is observed for pressure below a ``capillary'' pressure.
Conversely a very large slippage of the fluid on this composite
interface is found in this superhydrophobic state. Second, we
propose a macroscopic estimate of the effective slip length on the
basis of continuum hydrodynamics, in order to rationalize the
previous MD results. This calculation allows to estimate the
effect of a heterogeneous slip length pattern on the composite
interface. Comparison between the two approaches are in good
agreement at low pressure, but highlights   the role of the exact
shape of the liquid-vapor interface at higher pressure. These
results confirm that small variations in the roughness of a
surface can lead to huge differences in the slip effect.
On the basis of these results, we propose some guidelines to
design  highly slippery surfaces, motivated by potential
applications in microfluidics.
\end{abstract}

\pacs{83.80.Fg, 68.35.Np, 68.10c}

\date{\today}
\maketitle

\section{Introduction}
The question of the hydrodynamic boundary condition (h.b.c.) for
simple fluids has received much attention recently, and a number
of experimental studies  have addressed this problem~
\cite{Churaev84,watanabe99,Vinogradova99,Pit2000,Baudry2001,
Craig2001,Granick2001,cheng2002,Granick2002,tretheway2002,Bonaccurso2002,
lpmcn2002,Bonaccurso2003,VinoYabukov,Breuer03}. Usually, when a
simple liquid flows over a motionless solid wall, its velocity is
assumed to be equal to zero near the wall. This assumption is a
postulate of hydrodynamics. Although it is widely verified for
macroscopic flows of simple fluids, its validity at small scales
remains an open question. Wall slippage is usually described in
terms of an extrapolation length, usually denoted as the slip
length, \textit{b} : this corresponds to the distance between the
wall and the position at which the linear extrapolation of the
velocity profile vanishes. Analytical calculations and numerical
simulations have shown that two parameters mainly affect the
h.b.c.: the {\it roughness of the surface}~\cite{Richardson73} and
the {\it solid-liquid
interaction}~\cite{Barrat99a,Barrat99b,Thompson97}. While
roughness is usual\-ly expected to decrease slippage, hydrophobocity
of the solid surface has been found to enhance the slip length.
>From an experimental point of view, many studies have been carried
out recently but they do not allow to draw a simple and unique
conclusion concerning the effect of roughness and solid-liquid
interaction on slippage, as  a huge dispersion  in the results is
observed for seemingly similar systems. For instance, on
``smooth'' hydrophobic surfaces, slip lengths $b$ were obtained
that range from a few nanometers \cite{Churaev84} up to a few
micrometers \cite{tretheway2002}. Even at a qualitative level,
some results appear to contradict each other. To cite an
intriguing point as an exemple, roughness has be found either to
increase~\cite{Granick2002} or to decrease the slip
length~\cite{Bonaccurso2002}. It is therefore important to clarify
this question from a fundamental point of view, in order to
understand the origin of this important variability of results.
>From a more ``practical'' perspective, the possibility for a
liquid to slip over a solid surface can also have  important
consequences for flows in microchannels. A large slip length
reduces dissipation at the surfaces and therefore affects
the pressure drop in a given channel geo\-me\-try. 
It also  may help  in reducing
 the hydrodynamic dispersion due
to the velocity gradient between the center and the channel
boundary, with important applications for exemple in chemical or
biological techniques of ana\-ly\-sis (micro-chromatography). Finally,
slippage is also an important phenomenon for flows in mesoporous
media, where the understanding of the effect of mechanical
dissipation is essential for applications such as
dampers~\cite{lefevre}.

In previous work~\cite{naturemat}, we have shown by Molecular
Dynamics (MD) simulations that the slip length is strongly
affected by the roughness of the substrate for non wetting
solid-fluid interactions,
a point which could explain some of the experimental
discrepancies. In particular, we have found, for a non wetting
patterned surface, a thermodynamic transition between two
different surface states as a function of the pressure : a
``normal'' state where the liquid occupies all the available
volume, and a ``super-hydrophobic'' state, where the liquid flows
over the roughness. In the ``normal'' state, the h.b.c. is well
described by a no slip condition (the roughness decreases the slip
length), whereas the ``super-hydrophobic'' state enhances a slip
h.b.c., with slip lengths greater than in the smooth case (the
roughness increases the slip length).


Our aim in this paper is twofold. First, using MD si\-mu\-la\-tions, we
present  a more detailed study of the transition between the
``normal'' and the ``super-hydrophobic'' state,
as a function of the roughness parameters, geo\-me\-try and pressure. 
Note that the geometry studied here consists of grooves with
different widths and heights. This provides an ideal geometry,
from which one could expect to extract simple rules. Slippage over
the surface is conversely studied in the different thermodynamic
states. Second, we try to rationalize the effect of the surface
heterogeneity on the measured slip length, as observed in the
previous MD simulations. To this end, we develop a macroscopic
description of the hydrodynamic flow on a microscopically
heterogeneous surface, characterized by a spatially varying h.b.c.
Beyond the com\-pa\-ri\-son to the MD results, this analysis allows us to
propose an optimized surface heterogeneity, designed
to obtain large slippage at the interface.

\section{Molecular Dynamics approach}

\subsection{Phase properties of heterogeneous surfaces}
We consider in this study a liquid confined between two parallel
solid walls. The bottom wall is decorated with a periodic array of
grooves of height $h$ and width $a$. Note that this differs from
the system considered in reference \cite{naturemat}, where a
periodic array of square dots had been considered. Periodic
boundary conditions are used in the directions parallel to the
walls, the length of a cell of simulation is $L$. In our
simulations, all the interactions are of the Lennard-Jones type~:
\begin{equation} \label{equLJ}
v_{ij}=4\epsilon\left[\left(\frac{\sigma}{r}\right)^{12}-c_{ij}\left(\frac{\sigma}{r}\right)^6\right]
\end{equation}
The fluid  and the solid atoms have the same molecular diameter
$\sigma$ and interaction energies $\epsilon$. The variation in the
$c_{ij}$ (the index $i,j=F,S$ refers to the fluid or solid phase)
is a convenient  control parameter that can be varied to adjust
the surface tensions. The solid substrate is described by atoms
fixed on a the (100) plane of an FCC lattice. We have worked with
two values of  $c_{FS}$, $c_{FS}=0.5$ or $c_{FS}=0.8$, which
correspond to  contact angles (deduced from  Young's law) of
$\theta=137^{o}$ or $\theta=110^{o}$,
respectively~\cite{Barrat99b}. The simulations are carried out at
constant temperature $k_{B}T/\epsilon$=1. In flow experiments,
 the
velocity component in the direction orthogonal to the flow was
thermostatted~\cite{Barrat99b}, in order to avoid viscous heating
within the fluid slab.

\begin{figure}
\begin{center}
\includegraphics*[width=8cm]{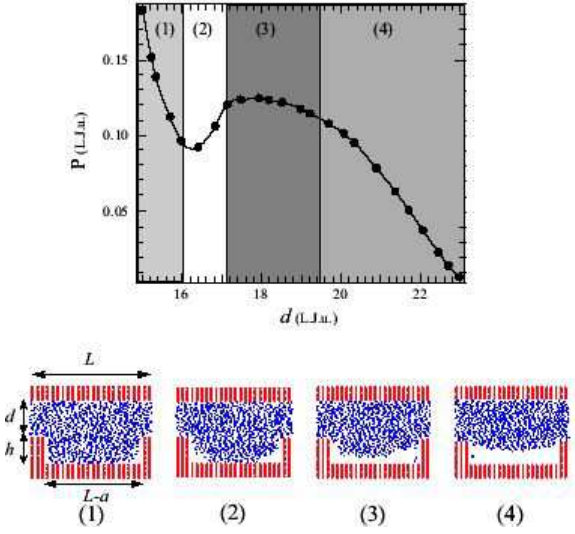}
\caption{(a) Normal pressure P (in Lennard-Jones units) versus
thickness of the cell $d$ (in Lennard-Jones units). The
periodicity is $L=18.9\ \sigma$, the height of the grooves is
$h=10.7\ \sigma$ and the width of the groove is $L-a=15.6\
\sigma$. (b) Transverse views of the atomic configuration for
different situations. Atoms belonging to the liquid and solid are
represented by points and round dots, respectively. }
\label{Pfoncdist}
\end{center}
\end{figure}

Figure~\ref{Pfoncdist} displays the evolution of the pressure with
the distance $d$ between the walls of the cell. The normal
pressure is obtained from the average force on the substrates
along the $z$ direction. Since the number of atoms  is kept
constant, this plot is similar to a $P-V$ isotherm (pressure
versus volume) in the bulk. We have verified that changing the
pressure by modifying the number of particules in the cell at
fixed volume $V$ leads to equi\-va\-lent results. As for the case of
dots \cite{naturemat}, this isotherm exhibits two branches,
separated by a ``Van der Waals loop'', indicating a phase
transition between two possible situations. As shown in figure
\ref{Pfoncdist}, where transverse views of the atomic
configurations
for different pressures 
are represented, this transition corresponds to a partial
dewetting of the space between the grooves. At  high pressures,
the liquid occupies all the cell (see figure \ref{Pfoncdist}),
including the grooves. This state will be described in the
following as the ``normal'' state. At lower pressures, partial
dewetting is observed and a composite interface is formed. In the
latter case, the space between the grooves is essentially free
from liquid atoms (see figure \ref{Pfoncdist}). This state will be
therefore denoted as the ``super-hydrophobic'' state. Note that
the range of distances for which the pressure decreases with $d$
is thermodynamically unstable, its observation being merely an
artefact of small size simulations.

Qualitatively, the maximum of the pressure on the previous Van der
Waals loop corresponds to the capillary pressure $P_{cap}$, for
which the liquid starts invading the grooves. On the other hand,
the minimum corresponds to the pressure at which the liquid
reaches the bottom of the groove and adapts its contact angle in
order to satisfy Laplace's law.

\begin{figure}
\begin{center}
\includegraphics*[width=6cm]{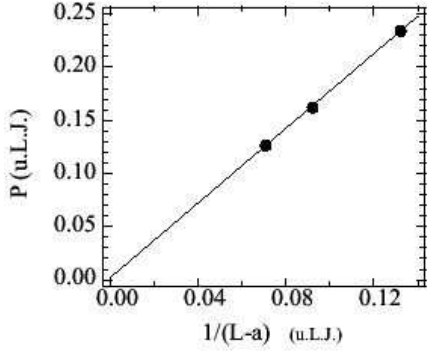}
\caption{Measured capillary pressure $P_{cap}$ in the simulation
(in Lennard-Jones units) as a function of the inverse groove width $1/(L-a)$.
These results corresponds to grooves with sufficiently large depth, $h$, so that
the measured capillary pressure becomes independent of $h$.
}
 \label{PLaplace}
\end{center}
\end{figure}

A qualitative
interpretation of this dewetting transition can be given in
terms of macroscopic capillarity. If we consider a system at fixed
normal pressure $P_ N$, the difference in Gibbs free energies
between the normal state (case $A$) and the formation of a
superhydrophobic state (case $B$) can be written as:
\begin{equation}\label{Fdem}
G_B-G_A=L \{(L-a+2h)\ (\gamma_{LV}cos\theta)+(L-a) \ \gamma_{LV}+P_N(L-a)h\}
\end{equation}
The composite interface is therefore favored when
\begin{equation}
P_N<P_{composite}=P_{cap}+\frac{-\gamma_{LV}.(cos\theta+1)}{h}
\label{pcoex}
\end{equation}
where we have introduced the capillary pressure $P_{cap}$, defined
as
\begin{equation}\label{Pcap}
P_{cap}=-\frac{2\gamma_{LV}cos\theta}{L-a}
\end{equation}
Although a quantitative agreement can hardly be expected in view
of the small sizes in our system, we have checked that equation
\ref{pcoex} correctly predicts the general trends observed in our
simulation, when the height or width or the grooves are varied. As
an example, we have plotted the measured capillary pressure on
Fig. \ref{PLaplace}, here defined in the simulations as the
maximum pressure for which the dewetted state is stable
(corresponding to the local maximum in figure  \ref{Pfoncdist}).
This plot corresponds to the limit in which the grooves have a
large depth $h$, so that the ``composite'' pressure, defined in
equation (\ref{pcoex}) should reduce to the capillary pressure,
equation (\ref{Pcap}). As shown in this figure, the linear
dependence of the  pressure as a function of $1/(L-a)$ does
confirm the validity of the  macroscopic argument above. Moreover,
the slope of the linear fit agrees within $15\%
$ with the
predicted one, $2\gamma_{LV}cos\theta$ \cite{Barrat99b}.

\subsection{Slippage over heterogeneous surfaces}

We now turn to the main point of the present paper, namely the
slippage of the fluid on the previous heterogeneous surface.

We have conducted parallel Couette flow simulations. A shear flow
is induced by moving the upper wall with velocity $U$ and the
lower wall with a velocity $-U$. Typically we chose $U=0.3$ in
reduced Lennard-Jones units. We have verified that, in spite of
the large value of the ve\-lo\-ci\-ty, the  response is linear and the
measured slip lengths are independent of the velocity $U$.
Temperature is kept constant by thermostatting through velocity
rescaling in the direction perpendicular to the flow. For flat
walls \cite{Barrat99a}, the slip length $b$ is defined as the
distance between the wall position and the depth at which the
extrapolated velocity profile reaches the nominal wall velocity,
$v=U$. In the presence of grooves, the same definition is used.
Obviously the presence of grooves makes the choice of the wall
position somewhat arbitrary. We choose to define the wall as the
position of the bottom layer of the substrate. This choice will be
seen to have only a negligible influence on the results (since, as
discussed below, the measured slip lengths are found to be much
larger than the groove depth in the interesting
``superhydrophobic'' state).

\begin{figure}[t]
\begin{center}
\includegraphics*[width=8cm]{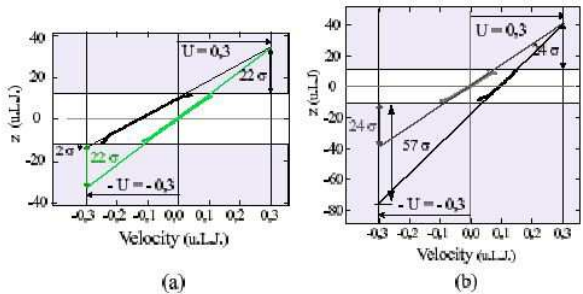}
\caption{Velocity profile in a Couette flow (black lines) : (a) in
the normal state (high pressure) where the liquid fills the
grooves; (b) in the superhydrophobic state (low pressure) where
the grooves are dewetted. The upper  and lower wall have velocity
$U=\pm 0.3$ (in Lennard-Jones units). The grey lines are the
velocity profiles obtained for a smooth surface ({\it i.e.} no
grooves) at the same pressure.
} \label{velocity}
\end{center}
\end{figure}

Typical measured velocity profiles are plotted in figure \ref{velocity}. 
As shown on this figure, the slippage strongly depends on the
surface state : globally, while in the ``normal'' state the slip
length is found to be very small (roughness reduces slip), in the
``superhydrophobic'' state the slip length is strongly enhanced
compared to the smooth surfaces case (with the same solid-fluid
interactions). One may note that for the smooth  top surface the
slip length remains almost constant with the pressure, and its
evolution is not affected by the transition between the ``normal''
and the ``super-hydrophobic'' state. This observation shows that
for the confinement we consider here, the dynamics in the vicinity
of the top wall is not affected by the bottom wall.

\begin{figure}[b]
\begin{center}
\includegraphics*[width=8cm]{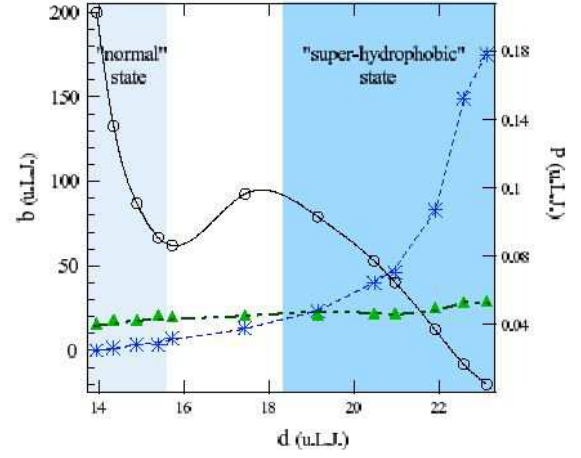}
\caption{Evolution of the slip length $b$ as a function of the
distance $d$ between the walls. The triangles ($\triangle$)
represent the values of the slip lengths on the smooth wall, the
stars ($\ast$) represent the  values of the slip lengths on the
rough wall, characterized by periodic grooves of height $h=10.7\
\sigma$, width $a=5\ \sigma$, periodicity $L=19.7\ \sigma$
and $c_{FS}=0.5$ (Lennard-Jones units). The flow is parallel to the grooves. The
circles ($\circ$) indicate the value of the pressure (scale on the
right of the figure). The lines are guides for the eye.}
\label{lpalisseh}
\end{center}
\end{figure}

We present in figure~\ref{lpalisseh} the evolution of the slip
lengths on the smooth wall and on the rough wall, with the
distance between the walls, together with the measured pressure
(here the number of molecules per cell is fixed). 

The results in figure~\ref{lpalisseh} were obtained for a {\it
velocity pa\-ral\-lel to the grooves}. We have performed the same
analysis for a velocity {\it perpendicular to the grooves} and
obtained very similar results, which typically differ by a
numerical constant (the slip in the direction perpendicular to the
grooves being less pronounced  than in the direction parallel to
the grooves, as expected). In order to simplify the discussion, we
will only present in the following the results for a liquid
flowing parallel to the grooves.

As mentioned above, in the normal state, the roughness is found to
reduce the slip. The slip length in the ``normal'' state is lower
on the rough wall than on the smooth wall. On the other hand, in
the ``super-hydrophobic'' state, roughness considerably enhances
the slip length. In the ``super-hydrophobic'' state, one observes
a strong dependence of the slip length with the pressure, an
effect which can be correlated to the strong dependence of the
shape of the liquid/vapor interface with the pressure. Indeed, as
the pressure decreases, the interface changes from an incurvated
shape in the grooves to a flat shape laying on the top of the
roughness. \\

\begin{figure}[b]
\begin{center}
\includegraphics*[width=6cm]{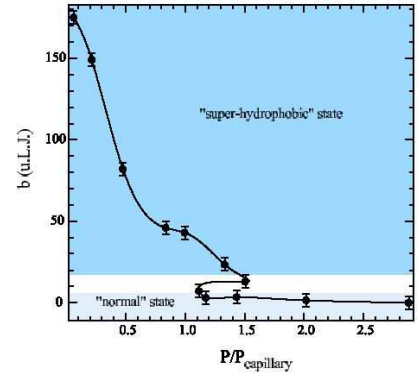}
\caption{Evolution of the slip length (in Lennard-Jones units) as
a function of $P/P_{capillary}$ for the same geometry as in
fi\-gu\-re~\ref{lpalisseh}.} \label{LP}
\end{center}
\end{figure}

In order to compare with experiments, it is more convenient to
plot the slip length as a function of  pressure (figure~\ref{LP}).
In this figure the pressure is made dimensionless using the
capillary pressure defined in equation (\ref{Pcap}).
>From this kind of representation one may expect to be able to
estimate the slip length for a specific roughness geometry,
knowing the pressure of the system. From this figure, it is
already obvious that small changes in the pressure induce very
large difference in the measured slip length.

\begin{figure}[t]
\begin{center}
\includegraphics*[height=6cm]{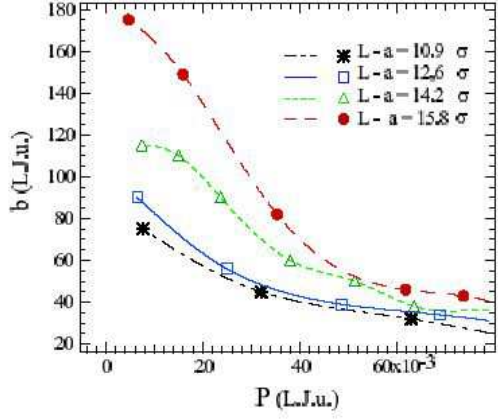}
\caption{Evolution of the slip length $b$ as a function of $P$
(Lennard-Jones units) for grooves of different width $L-a$ and
same height $h=9.6\ \sigma$. We only represent the slip lengths in
the ``super-hydrophobic'' states.} \label{evolutionblargeur}
\end{center}
\end{figure}

We now turn to the influence of the geometry of the roughness on
the slip length. We mainly focus on the role of the groove's
width. Figure~\ref{evolutionblargeur} represents the evolution of
the slip length as a function of the pressure for various widths
of the grooves (keeping their height constant in this case). For a
given pressure the slip length increases with the width of the
grooves. This point can be ea\-si\-ly understood. A wider  groove
implies a larger area of  liquid/vapor interface, leading to a
lower friction at the interface. From this figure we also note
that small va\-ria\-tions in the geometry of the grooves can lead to
very large variations in the slip lengths. This fact could explain
the great variability observed in the experimental
results. 
Small differences in the state of the surfaces used in experiments
could lead to huge differences on the slip lengths.


\subsection{Conclusions and hydrodynamic description}

At this stage we can draw a few general conclusions. On the basis of MD simulations
we have shown that the surface roughness is a key factor in understanding
the slippage of fluids on (hydrophobic) solid surfaces. Depending on the applied
pressure a dewetting transition is found, which conversely enhances considerably
the slip length at the surface.

We now try to rationalize this effect at the level of continuum
hydrodynamics, in order to capture the dependence of the slip
lengths as a function of the pe\-rio\-di\-ci\-ty of the roughness. Indeed,
in the dewetted state the liquid/wall interface takes the form of
a succession of stripes, alternating a liquid-solid interface with
a liquid-vapor interface. At the most simple level, the surface
can be  modelled as stripes with different slip lengths. To this
end, we develop a macroscopic description of the hydrodynamic flow
on a microscopically heterogeneous surface, characterized by a
spatially varying slip length.

\section{Macroscopic approach : flow over a surface with a
heterogeneous microscopic boundary condition}

\subsection{Introduction}

We study here the influence of a microscopic heterogeneous h.b.c.
on the macroscopic scale. Our aim is to define a {\it macroscopic slip length}
corresponding to a he\-te\-ro\-ge\-neous microscopic h.b.c., with a spatially varying value
of the slip length.
We therefore consider a semi-infinite situation, where a shear flow, with
a given shear rate $\dot\gamma$, is imposed at infinity.
Using hydrodynamics, we show that far from the heterogeneous interface,
that is to say for distances larger than the characteristic size of the structure,
an {\it effective slip length} can be defined, which depends on the underlying
``microscopic'' slip he\-te\-ro\-ge\-nei\-ty.

We model the composite surface by a planar surface ($Oxy$ plane)
characterized by a local slip length $b(x,y)$ function of  the in
plane coordinates $(x,y)$. Relying on the microscopic results of
Molecular Dynamics, one may assume for instance that $b$ is
locally very high over a vapor bubble and much lower elsewhere
(for simple mo\-dels, we may consider respectively infinite and zero
slip lengths). The characteristics of the flow far away from the
surface can then be determined by solving the hydrodynamic
equations with the h.b.c. locally given by $b(x,y)$. Such an
approach has been previously used to study local slipping
heterogeneities of a simple geo\-me\-tric shape : periodic stripes
parallel or perpendicular to the flow, corresponding to the
alternance of a no slip boundary condition and an infinite slip.
Philip has studied stripes parallel to the flow in a circular pipe
of radius $R$~\cite{Philip}. When $R\rightarrow\infty$, this
configuration is the same as considering a flow at infinite over a
plane, which is the situation studied here. In this limit, he
obtains for the macroscopic slip length
\begin{equation}
B_{//}=\frac{L}{\pi}\ \textrm{ln}\
\left(\frac{1}{\textrm{cos}\left(\zeta\frac{\pi}{2}\right)}\right)
\label{Philip}
\end{equation}
where $L$ is the periodicity of the pattern and $\zeta$ is the
fraction of the surface where the slip length is infinite.

Lauga and Stone have established the expression of the macroscopic
slip length for similar conditions than Philip (flow in a pipe of
radius $R$ with $\zeta$ the fraction of infinite slip), but with a
 pattern perpendicular to the flow~\cite{lauga}.  When
$R\rightarrow\infty$, their expression for the macroscopic slip
length reduces to
\begin{equation}
B_{\bot}=\ \frac{L}{2\pi}\ \textrm{ln}\
\left(\frac{1}{\textrm{cos}\left(\zeta\frac{\pi}{2}\right)}\right)\
.
\end{equation}

We are interested here in a more general situation, involving  an
arbitrary pattern (both in geometry and ma\-gni\-tu\-de)   of the
 microscopic slip length.

\subsection{Geometry and hydrodynamic equations}

\begin{figure}[b]
\begin{center}
\includegraphics*[height=4cm]{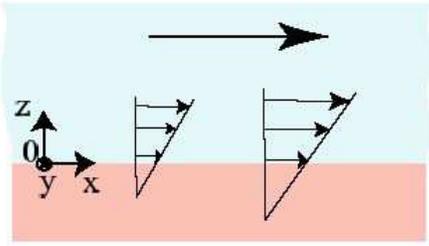}
\caption{Schematic representation of the velocity field of the
liquid for $z>0$, for a flow in the $x$ direction.}
\label{ecmacro}
\end{center}
\end{figure}

We consider an incompressible simple liquid occupying the
half-space $z>0$, with an imposed shear stress $\dot{\gamma}$ in
the $x$ direction and flowing over a fixed solid plane located in
$z=0$ as shown in figure~\ref{ecmacro}.

We now consider a stationary regime at low Reynolds number so that
the Navier-Stokes equations can be written as
\begin{equation}
\left\{
\begin{array}{rl}
\eta  {\bf \nabla}^2 \bf{U} & =\bf{\nabla}\ P\\
\bf{\nabla} \cdot(\bf{U}) & =0
\end{array}
 \right. ,
 \label{eq:3}
\end{equation}
where $\bf{U}$ is the velocity vector, $P$ is the pressure and
$\eta$ the viscosity of the liquid. The boundary
conditions are defined as follows :\\
\\
$\bullet$ \underline{at $z=0$} : a given local slip boundary
condition $b(x,y)$ which is a function of  the in plane
coordinate.
\begin{equation}
\left\{
\begin{array}{rl}
b(x,y)\left.\frac{\partial{U}_x}{\partial z}\right|_{z=0} & =U_x(x;y;z=0)\\
\\
b(x,y)\left.\frac{\partial U_y}{\partial z}\right|_{z=0} &
={U}_y(x;y;z=0)\\
\\
 {U}_z(x;y;z=0) & =0
\end{array}
 \right.
 \label{eq:hbc}
\end{equation}
\\
$\bullet$ \underline{for z $\rightarrow\infty$} :
\begin{equation}
{\partial{{\bf U}}\over{\partial z}}=\dot{\gamma}{\bf e}_x
\end{equation}

The effective slip length is defined in terms of the asymptotic
velocity field, ${\bf U}$, solution of the above equations. In the
$z\rightarrow \infty$ limit, one expects
\begin{equation}
{\bf U}={\bf U}_S
+\dot{\gamma}z\bf{e}_x=(\alpha+\dot{\gamma}z)\bf{e}_x+\beta{\bf
e}_y,
\end{equation}
where ${\bf U}_S$ defines the ``effective'' velocity slip on the
wall, with components \{$\alpha, \beta$\}: ${\bf U}_S$ is a
constant velocity vector (independent of $x$ and $y$), parallel to
the wall, which defines the \textit{macroscopic} slip vector at
the wall.
The hydrodynamic equations~(\ref{eq:3}) and the boundary
conditions being linear, ${\bf U}_S$ must be proportional to
$\dot{\gamma}$. The macroscopic slip lengths are then defined as
$B_x=\alpha/\dot{\gamma}$ and $B_y=\beta/\dot{\gamma}$, here
associated
with  a flow in the $x$ direction.\\
\\
To solve this problem we introduce  a new vector field
\begin{equation}
{\bf V}={\bf U}-\dot{\gamma} z {\bf e}_x \label{defV}
\end{equation}
and eliminate  the pressure by using the vorticity
\begin{equation}
{\bomega}={\nabla} \times \bf{V}
\end{equation}
The system~(\ref{eq:3}) can be rewritten as :
\begin{equation}
\left\{
\begin{array}{rl}
\nabla^2 \bomega & =\bf{0}\\
\bf{\nabla}\cdot ({\bf V}) & = 0
\end{array}
 \right.
\label{eq:laplome}
\end{equation}
We consider a  periodic pattern,    with
periodicity $L$ in both directions.  Dimensionless variables (starred in the notations) are defined as\\
 \noindent

\begin{equation}\label{eq:adimension}
\left.
\begin{array}{l}
\displaystyle{x=x^\ast \frac{L}{2\pi}}\ ;\ \displaystyle y=y^\ast
\frac{L}{2\pi} \ ;\  \displaystyle z=z^\ast
\frac{L}{2\pi} \ ;\\
\\
b=b\displaystyle ^\ast \frac{L}{2\pi} \ ;\\
\\
 V=V\displaystyle ^\ast \dot{\gamma} \frac{L}{2\pi} \ ;\
\displaystyle \omega = \omega^\ast \dot{\gamma} \ ;\\
\displaystyle \alpha = \alpha^\ast \dot{\gamma} \frac{L}{2\pi} \
;\ \displaystyle \beta=\beta^\ast \dot{\gamma} \frac{L}{2\pi} ;
\end{array}
\right|
\end{equation}

To simplify notations, we will drop  the stars on the variables in
the following, and all length and velocity variables must be
understood as  being rescaled by the periodicity of the pattern.
The description of the re\-so\-lu\-tion of equations~(\ref{eq:laplome})
with h.b.c.~(\ref{eq:hbc}) is given in the appendix, where we show
that the system can be conveniently
rewritten in terms of the 2D  Fourier components
of the various fields in the $x,y$ plane (we do not consider the Fourier transform in the $z$ direction):\\
For $\vec q\neq0$ :
\begin{equation}
\left\{
\begin{array}{rl}
\left[\hat{b}\circ\underline{\hat{\omega}}_y\right]({\bf
q})+\hat{b}({\bf q}) &=\frac{q_x}{2q^3}\left(q_x
\underline{\hat{\omega}}_y({\bf q})-q_y\underline{\hat{\omega}}_x({\bf q})\right)-\frac{\underline{\hat{\omega}}_y({\bf q})}{q}\\
-\left[\hat{b}\circ\underline{\hat{\omega}}_x\right]({\bf
q}) & =\frac{q_y}{2q^3}\left(q_x
\underline{\hat{\omega}}_y({\bf
q})-q_y\underline{\hat{\omega}}_x({\bf
q})\right)+\frac{\underline{\hat{\omega}}_x({\bf q})}{q}
\end{array}
\right. \label{eq:sysnis1}
\end{equation}
For $q=0$ and $z=0$ :
\begin{equation}
\left\{
\begin{array}{rl}
\left[\hat{b}\circ\hat{\omega}_y\right]({\bf q}={\bf
0})+\hat{b}({\bf q}={\bf 0})
& = \alpha \\
-\left[\hat{b} \circ \hat{\omega}_x\right]({\bf q}={\bf 0}) & = \beta \\
\end{array}
 \right.
 \label{eq:sol}
\end{equation}
where the notation ``$\circ$'' stands for the convolution pro\-duct.

The previous system  for $\underline{\hat{\omega}}_x({\bf q})$ and
$\underline{\hat{\omega}}_y({\bf q})$, in equation
(\ref{eq:sysnis1}), is solved numerically via a matrix inversion.
The macroscopic slip lengths, $B_x$ and $B_y$, defined in terms of
$\alpha$ and $\beta$, are then deduced by substituting this
solution into (\ref{eq:sol}). Calculation details are given in the
appendix.

In the following section we present the results obtained for
various patterns of the microscopic hydrodynamic boundary
condition.

\subsection{Alternating stripes with perfect slip and no-slip h.b.c. }\label{sec:slipinf}

\begin{figure}
\begin{center}
\includegraphics*[height=4cm]{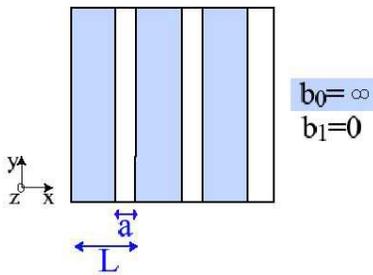}
\caption{Schematic representation of the microscopic hydrodynamic
boundary condition, alternance of strips of no-slip and perfect
slip.} \label{altbandes}
\end{center}
\end{figure}

As sketched in figure~\ref{altbandes}, we consider a succession of
stripes of width $a$ characterized by a vanishing slip length
($b_1=0$), and stripes of width $L-a$ with an infinite slip length
($b_0 \rightarrow \infty$). The fraction of the surface  with
slippage is $\zeta=(L-a)/L$. We can consider stripes in any
direction with respect to the flow, but we will mainly focus on
the case of stripes parallel or perpendicular to the flow. This
system corresponds obviously to a ``model'' experimental situation
corresponding to a succession of grooves with vapor bubbles
(slipping stripes) and rough strips (non slipping stripes). We
present below the results for the effective macroscopic slip
length $B$ in this situation.

\begin{figure}
\begin{center}
\includegraphics*[height=12cm]{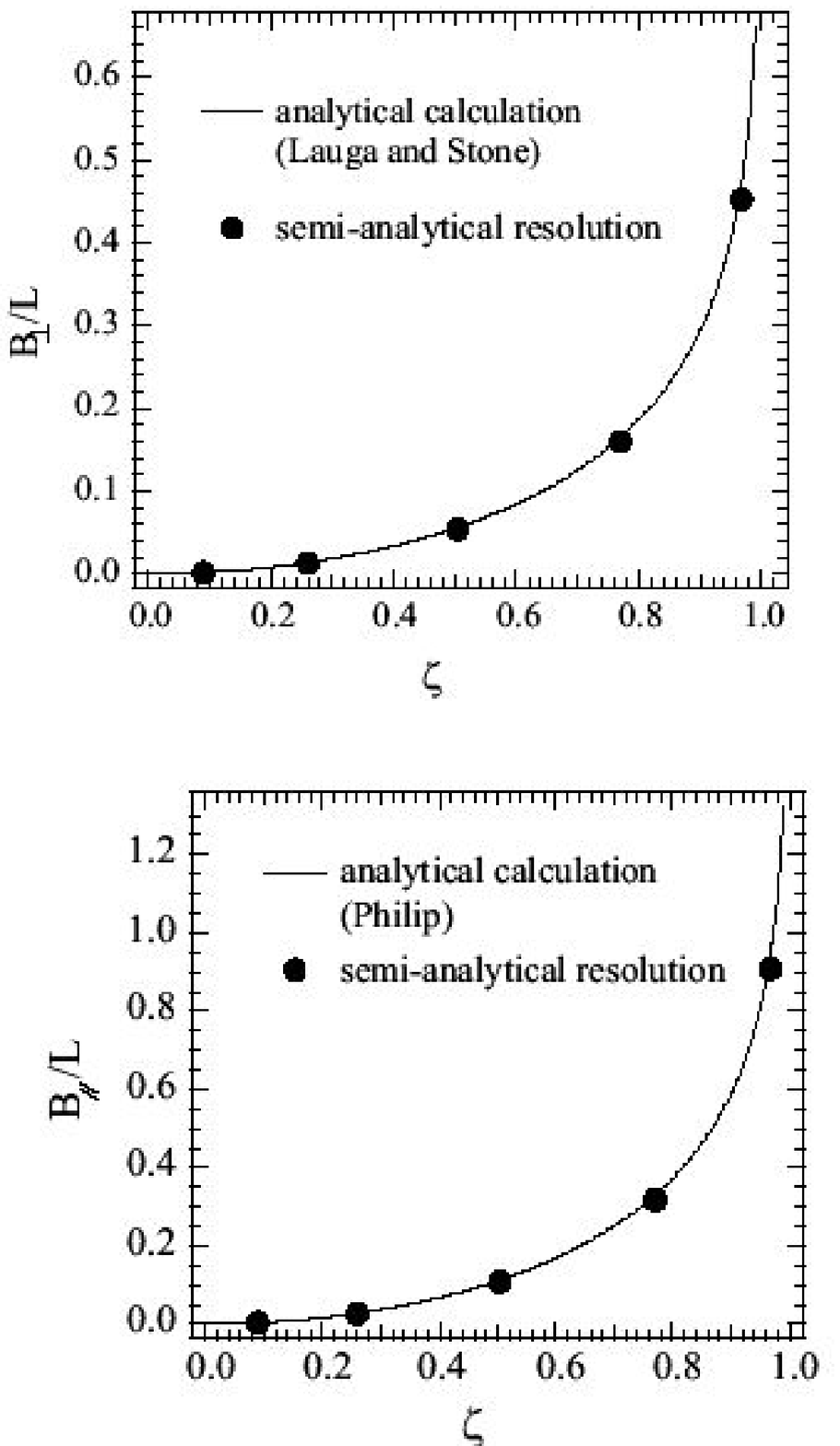}
\caption{We consider an succession of stripes of width $a$ where
the microscopic slip length is equal to zero ($b_1=0$) and strips
of width $L-a$ with an infinite microscopic slip length ($b_0
\rightarrow \infty$) and represent the evolution of the
macroscopic slip length $B$ as a function of $\zeta$ ($\zeta$ is
the fraction of the surface with a microscopic infinite slippage,
$\zeta=(L-a)/L$). The curves represent the analytical calculations
and the dots the numerical resolution (top: flow perpendicular to
the strips, bottom : flow parallel to the strips.)}
\label{analmacro}
\end{center}
\end{figure}

First we study the influence of the
fraction of the slipping surface, $\zeta$, on the macroscopic slip length .
This case is a benchmark for our calculation since in this
specific situation we can compare our results with those obtained
by Philip \cite{Philip} and Stone and Lauga \cite{lauga} in the
limit of a cylinder with an infinite radius $R\ \rightarrow\
\infty$~\cite{lauga,Philip}. As shown on figure~\ref{analmacro},
our approach is in excellent agreement with their analytical
results. This validates  the semi-analytical approach developed
here.

An interesting result emerging from the calculation in this
specific geometry, is that a small percentage of non slipping
surface ($1-\zeta \sim 0$) is enough to decrease
con\-si\-de\-ra\-bly the effective slip~: 
even for $\zeta=98.5 \%
$,  $B$ is only a slightly larger than the size of the
pattern of he\-te\-ro\-ge\-nei\-ty $L$ (here with a shear
parallel to the pattern).

\subsection{Alternating stripes of no-slip and partial slip h.b.c.}

\begin{figure}[htbp]
\begin{center}
\includegraphics*[height=14cm]{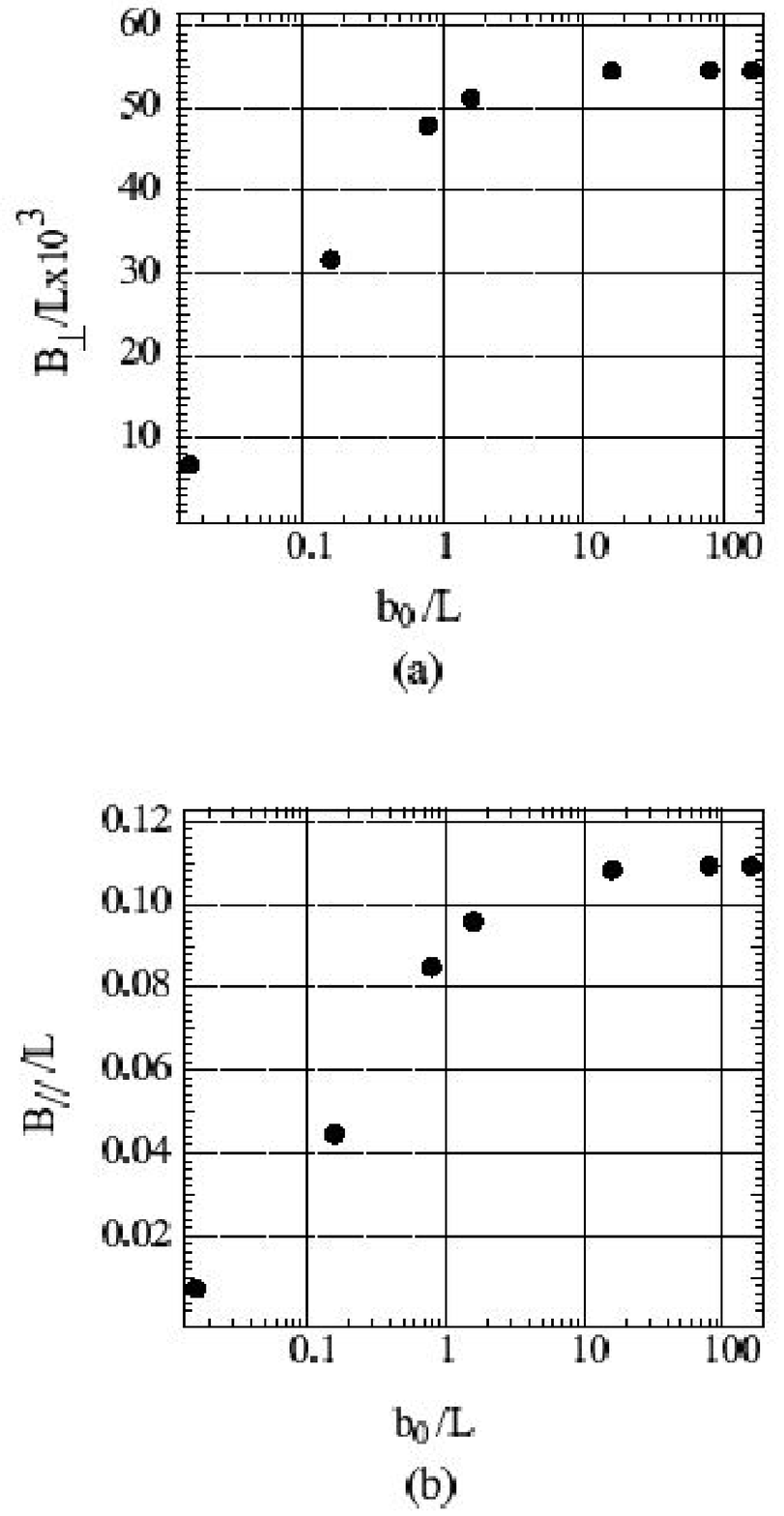}
\caption{Macroscopic slip length $B$ on a surface composed of alternating stripes with
slip length $b_0$ and vanishing slip length, 
for $\zeta=50\%
$. On figure (a), the shear flow is
perpendicular to the slipping stripes; on figure (b), the shear flow is
parallel to the slipping stripes
(note the change of vertical scale).} \label{fig:Bcpab}
\end{center}
\end{figure}

We now consider a different situation, with a surface composed of
alternating non slipping stripes ($b_1=0$), and stripes with partial
slip h.b.c., characterized by a slip length $b_0$.  The surface fraction
with slip length $b_0$ is again defined as $\zeta$.
Our semi-analytical approach allows to consider any value of
$b_0$, and extends therefore the  results available  in the
literature, which are restricted to the $b_0=\infty$ case. We
first present the results for $\zeta=50\%
$. We have plotted in
figure~\ref{fig:Bcpab} the evolution of the macroscopic slip
length as a function the microscopic slip length $b_0$ in the case
of a flow parallel or perpendicular to the strip pattern. The main
observations are  :
\begin{itemize}
\item For  low values of $b_0$ ($b_0<L$), the macroscopic slip
length $B$ increases linearly with $b_0$ (of the order $b_0/4$ for
the percentage of the slipping surface considered). \item For
high values of $b_0$, ($b_0\ >\ 10L$) B asymptotically tends to a
limiting value that is a fraction of the periodicity (around
$L/10$ for a flow parallel to the strips and $L/20$ for a flow
perpendicular to the strips).
\end{itemize}

\begin{figure}[htbp]
\begin{center}
\vspace{5cm}
\includegraphics*[height=6cm]{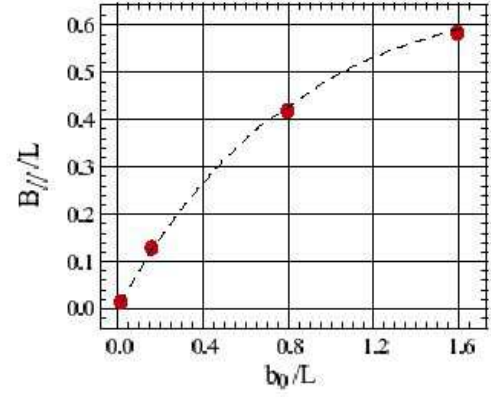}
\caption{Macroscopic slip length $B$ on a surface composed of alternating stripes with
slip length $b_0$ and vanishing slip length, as a
function of the microscopic slip length $b_0$, for $\zeta=97\%
$. The shear flow is parallel
the slipping strips.} \label{fig:Bpa97}
\end{center}
\end{figure}

The important point is that in the present $b_1=0$ case,
\textbf{the
macroscopic slip length $B$ is fixed by the smallest of the two lengths
$b_0$ and $L$}.\\
\\
The same conclusions hold for larger values of $\zeta$.
Figure~\ref{fig:Bpa97} shows the results obtained for the
effective slip length in the direction parallel to the stripe,
$B_\parallel$, for a surface with a fraction $\zeta=97\%
$
characterized by a slip length $b_0$. Due to numerical
limitations, we could only study   slip lengths $b_0/L<2$.  The
results are similar to those obtained at $\zeta=50\%
$.

\subsection{Alternating stripes of infinite slip and partial slip h.b.c.}

We now consider a succession of stripes with an infinite slip
length ($b_0\ \rightarrow\ \infty$) and stripes with a finite slip
length $b_1$ (now with a non vanishing $b_1$). Obviously, this
configuration is expected to model the situation considered in the
first part of this work, specifically in the ``super-hydrophobic''
state. We have shown in the first part of this work that the
interface between the liquid and the rough surface takes, in this
state, the form of a succession of liquid-solid stripes,
characterized by a finite slip length ($b_1\sim 20 \sigma$, with
$\sigma$ the molecular diameter), and liquid-vapor stripes, {\it a
priori} characterized by an infinite slip length. We come back to
this assumption in the following.

\begin{figure}[htbp]
\begin{center}
\includegraphics*[height=6cm]{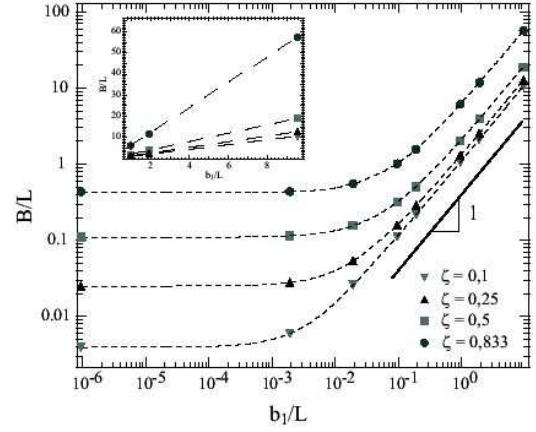}
\caption{Macroscopic slip length $B$ on a surface composed of alternating stripes with
slip length $b_1$ and infinite slip length, as a
function of the microscopic slip length $b_1$. $B$ is computed for a shear parallel
the slipping strips 
and for different values of $\zeta$. Dotted lines are a guide for the eye. In the inset the same points
are plotted on a linear scale. The dashed lines are linear fits
to the measured values of $B$.} \label{fig:Bb1zoom}
\end{center}
\end{figure}

Figure~\ref{fig:Bb1zoom} shows the evolution of the macroscopic slip
length $B$ as a function of the microscopic slip length $b_1$ for
different fractions $\zeta$ of the slipping surface  ($\zeta$ is defined
here as the fraction of the surface with an infinite microscopic
slip length, the rest of the surface having a finite slip length
$b_1$).
As shown on this figure, 
For $b_1> L $, the effective slip length $B$ evolves linearly with $b_1$ for all the fractions
$\zeta$.
On the other hand, one finds that for $b_1=0$ or $b_1 \ll L$ the macroscopic slip
length is fixed by the heterogeneity's periodicity $L$, in
agreement with the result of the section~\ref{sec:slipinf} (for
$b_1=0$ and $b_0\rightarrow\infty$).
The important conclusion we can draw from these results is
that in this geometry, {\bf with
stripes of infinite slip length and finite slip length,
the value of the macroscopic slip length is
determined by the largest of the finite lengths : $b_1$ or $L$}.

A simple phenomenological model 
can be formulated
to explain the linearity with $b_1$ in the case $b_1>L$. We introduce
the interfacial friction coefficient $\lambda$, defined
by writing the continuity of the tangential stress $\sigma_s$ at
the solid-liquid interface :
$$\sigma_s= \eta \,\frac{\partial\textrm{V}}{\partial z}\ =\ \lambda \textrm{V}_s$$
where $\eta$ is the viscosity of the liquid and V$_s$ the velocity
slip. The interfacial friction $\lambda$ coefficient is then
related to the slip length $b$ by:
$$\lambda=\frac{\eta}{b}$$
The effective friction coefficient $\Lambda=\frac{\eta}{B}$ can be
interpreted as the {\it averaged friction} over the different
strips, and we obtain accordingly the following result for the
effective macroscopic slip length as a function of the microscopic
ones~:
\begin{equation}
\frac{1}{B}=\frac{\zeta}{b_0}+\frac{1-\zeta}{b_1} \ ,
\label{Beff}
\end{equation}
which  is similar to the addition rule for resistors in pa\-ral\-lel.

In the case $b_0\ \rightarrow\ \infty$, we expect :
\begin{equation}
B=\frac{b_1}{1-\zeta} \label{eqBb}
\end{equation}
We have checked the linear dependance of $B$ with $b_1$ as well as
the dependance with $\zeta$ (table~\ref{tablezeta}) and we find
that the relation~\ref{eqBb} is very well verified for $b_1>L$.
It is important to emphasize that its validity is limited to the
case where both the slip lengths, $b_0$ and $b_1$ are larger than
the roughness periodicity $L$. 
Note however that in practice, this relationship is valid down to $b_1>0.1 L$ (see figure \ref{fig:Bb1zoom}).

For $b_1\ll L$ or $b_1=0$ the
friction coefficient tends to diverge over the strip and the model
does not apply. Physically one may attribute this failure in this
limit to the fact that the flow  is strongly perturbed (compared
to the asymptotic form) on the poorly slipping  stripes ($b_1 \ll
L$). This  is not taken into account in the simple argument
leading to Eq. (\ref{eqBb}).

\subsection{Comparison to Molecular Dynamics results}

We are now in a position to compare the hydrodynamic predictions
to the MD results.
We focus on the super-hydrophobic state where the measured slip length
is much larger than the bare slip length $b_1$ on the smooth surface.
In order to make the comparison with the macroscopic
approach, it is necessary to specify the local slip lengths on
the top of the roughness (solid-liquid interface) and at the
liquid-vapor interface on the grooves.

The analysis of the local profiles of velocity suggests an
infinite slip length at the liquid-vapor interface ($b_0=\infty$),
over a stripe of width $L-a$. For the slip length on the
solid-liquid interface, on the top of the roughness, we take the
value of the slip length $b_{smooth}$ obtained on a smooth wall at
the same pressure ($b_1=b_{smooth}$ over a strip of width $a$). We
only present the results in the case of shear parallel to the
grooves. As we have mentioned before the slip length of a confined
liquid depends on the pressure while there is no pressure
dependence in the macroscopic approach. To introduce the pressure
in the macroscopic approach we take into account the dependence of
$b_1=b_{smooth}$ with the pressure. This dependence is measured
independently in the simulations and depicted e.g. in figure
\ref{lpalisseh}.

\begin{figure}[htbp]
\begin{center}
\includegraphics*[height=7cm]{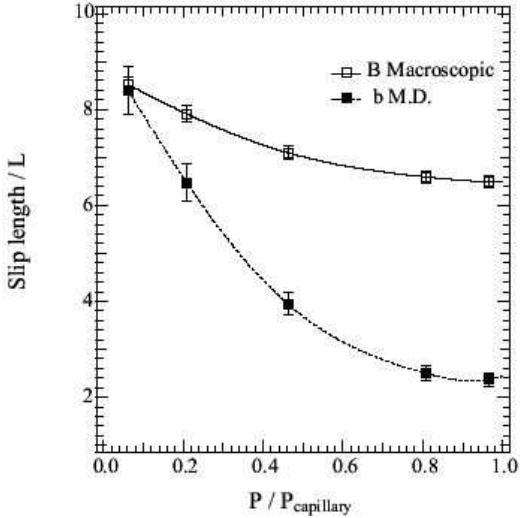}
\caption{Comparison between MD measures of the slip length in
the superhydrophobic state, and the macroscopic prediction of the slip length $B$ as a
 function of the pressure. The shear is
parallel the slipping stripes. 
The macroscopic estimate assumes alternating stripes with finite
slip length $b_1$ and infinite slip length; $b_1$ is accordingly
identified with $b_{smooth}(P)$ the pressure dependent slip length
on a smooth (and flat) solid surface, as measured independently in
MD simulations. The percentage  of the surface with an infinite
slip length (macroscopic simulation) is identified as the
percentage of the surface with a liquid-vapor interface (in MD
simulations) : here $\zeta=83.3 \% 
$. Lines are just a guide for
the eyes.} \label{fig:DMmacro}
\end{center}
\end{figure}

The macroscopic slip length B is then computed u\-sing the
macroscopic approach above, as a function of the pressure, in the
case of alternated stripes with infinite slip length
($b_0=\infty$) and stripes with slip length $b_1=b_{smooth}(P)$.
We compare these predictions  with those obtained by molecular
dynamic simulations with the same fraction $\zeta$ of surface with
an infinite slip length. In figure~\ref{fig:DMmacro}, we compare
 the slip length $b$ obtained in MD
simulations for various pressures with the macroscopic slip length
$B$ computed along the lines described above. The perfectly
slipping surface fraction is $\zeta=83.3\%
$. The pressure is made
dimensionless using the capillary pressure.

A few comments are in order.
\begin{itemize}
\item For low reduced pressures, $P/P_{cap} < .2$, there is a good
and even quantitative agreement between the MD results and
hydrodynamic predictions. This suggests that the hydrodynamic
prediction, with a surface patterned with stripes characterized by
infinite slip lengths and finite slip lengths, does capture the
essential ingredients of heterogeneous slippage in the
super-hydrophobic state. \item Nevertheless we see from figure
\ref{fig:DMmacro} that this agreement deteriorates when the
pressure approaches  the capillary pressure, $P_{cap}$. In
particular, the dependence of the slip length on the pressure is
more pronounced in the case of the molecular dynamics simulations
than in the macroscopic approach.
\end{itemize}

\begin{figure}
\begin{center}
\includegraphics*[width=6cm]{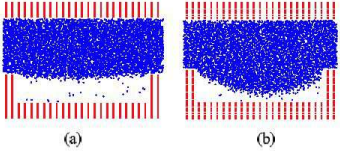}
\caption{Transverse views of the atomic configuration for the same
geometry as the one studied on figure~\ref{fig:DMmacro} for (a)
P/P$_\textrm{capillary}$=0.4 and (b)
P/P$_\textrm{capillary}$=0.9.} \label{fig:micrmacr}
\end{center}
\end{figure}

A possible explanation for this 
difference  is related to the geometry of the liquid-vapor
interface which evolves as the pressure reaches $P_{cap}$, an
ingredient which is not taken into account in the macroscopic
approach.  Figure \ref{fig:micrmacr} shows the evolution of the
shape of the liquid-vapor interface, as measured in MD
simulations. As can be seen from this figure,
this interface strongly depends on the pressure and 
takes a curved shape for pressure close to $P_{cap}$.
This suggests that the assumption  of a plane liquid-vapor interface in the macroscopic approach
in not appropriate at high pressure.  
The difference between the
MD and hydrodynamic results indicates that the curved surface induces a supplementary
dissipation as compared to the flat liquid-vapor interface, resulting in a lower effective
slip length. Indeed, at high pressure, the liquid penetrates in the groove so that
the flow is characterized by an increase of the velocity gradients compared to the low pressure case.
These velocity gradients lead to a supplementary dissipation as compared to the flat
interface, which could explain the difference between the two approaches at pressures
close to the capillary pressure.

To conclude  this section, this analysis points out the growing
importance of the shape of the interface on the global friction at
the interface, as pressure is increased.

\section{Conclusions and perspectives}

In this paper, we have considered the effect of surface heterogeneity on the
slippage of fluid. Two approaches have been followed. First, we have used MD simulations
to measure slip lengths on rough hydrophobic surfaces. A dewetting transition,
leading to a super-hydrophobic state,
is exhibited at small pressure, and leads to very large slippage of the fluid on this
composite interface. Our results therefore
confirm that mesoscopic roughness at the solid
liquid interface can drastically modify the interfacial flow
properties, in the same manner as it affects static wetting
properties.
Then on the basis of continuum hydrodynamics, we have proposed
a macroscopic estimate of the
effective slip length associated with a surface characterized by a heterogeneous
slip length pattern.
This  (semi-analytical) approach has enabled us to
determine the important factors controlling the effective slip length.
In particular we have characterized the effective slip length $B$
for a surface with a striped microscopic hydrodynamic boundary condition
characterized by slip lengths $b_0$ and $b_1$ with periodicity $L$.
A few ``rules of thumbs'' emerge from this analysis :
\begin{itemize}
\item for $b_1=0$, the effective slip length $B$
is controlled by the smallest of the two lengths,
$b_0$ or $L$; 
\item in the case where both slip lengths $b_0$ and $b_1$
are larger than the periodicity $L$, the effective slip length $B$ can
be approximated by a ``inverse law''
\begin{equation} 
\frac{1}{B}=\frac{\zeta}{b_0}+\frac{1-\zeta}{b_1}
\label{Beff2}
\end{equation}
with $\zeta$ the surface fraction with slip length $b_0$.
In practice, we found that this relationship holds down to $b_0, b_1 > {1\over 10} L$.
\end{itemize}
In this paper, we present results on the effective slip length $B$ associated
with a  surface formed by alternating stripes of different microscopic hydrodynamic boundary condition. However our results 
can be extended to more complex
slipping patterns, relevant to experimental systems.
As an example, surface patterns such as squares or dots can
be easily implemented within our semi-analytical approach.
Note that such microscopic patterns can not be solved in an analytical way.

We have then compared the results obtained within the two
approaches, MD results and hydrodynamic des\-crip\-tion.  While the
hydrodynamic approach reproduces quantitatively the MD results for
small pressures, a discrepancy between the two approaches is
exhibited at large pressures, close to the dewetting transition.
This points out the importance of the shape of the liquid-vapor
interface in the ``super-hydrophobic'' state on the dissipation at
the interface. The curvature of the liquid-vapor interface leads
to a supplementary dissipation on the liquid-vapor interface.

\begin{figure}
\begin{center}
\includegraphics*[height=5cm]{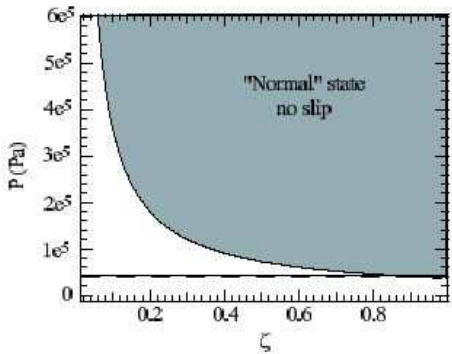}
\caption{Evolution of the capillary pressure as a function of
$\zeta$ ($P_{capillary}=\frac{-2\gamma_{lv}\cos{\theta}}{\zeta L}$) for
$\theta=105^\circ$and $\gamma_{lv}=72$~mN/m. In the grey zone the
pressure is higher than the capillary pressure, the liquid is in
the ``normal'' state, it is not possible to have high values of
the slip length. The dashed lines corresponds to the pressure drop
for microfluidic experiments. The intersection between this dashed
line and the capillary pressure's curve gives the upper value of
$\zeta$, here $\zeta=0.9$.} \label{fig:Pzeta}
\end{center}
\end{figure}

\begin{figure}[htbp]
\begin{center}
\includegraphics*[width=8cm]{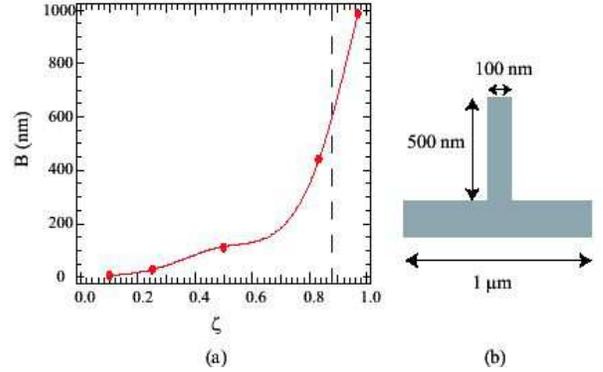}
\caption{(a) Evolution of the macroscopic slip length $B$ as a
function of $\zeta$, when $\zeta$~\% 
of the surface has an
infinite slip length and ($1-\zeta$)~\% 
of the surface has a slip
length of 20~nm (the periodicity is $L=1\mu$m). The vertical
dashed corresponds to the maximum value of $\zeta$ possible as
explained on figure~\ref{fig:Pzeta}. (b) Kind of structure that
can lead to slip lengths up to 600~nm.} \label{fig:Bzeta}
\end{center}
\end{figure}

Eventually, as an example of application of the previous
macroscopic and molecular dynamic approaches, we now apply these
results to design a potentially highly slipping surface. This is
motivated by potential applications, especially in microfluidics.
The motivation to design highly slipping surfaces is linked in
particular to the development of flat velocity profiles within
microchannel, aiming at reducing the
dispersion in the transport of different species. 
Such a flat profile can be obtained using a pressure drop flow
with a highly slipping channel. The studies we have presented show
that high slippage can be achieved using a hydrophobic channel
with ``appropriate'' roughness. We now have to design this
``appropriate'' roughness. For microfluidic applications, let us
assume a periodicity $L=1\mu m$. We have to determine the
percentage of the surface $\zeta=\frac{L-a}{L}$ which has an
infinite slip length. We assume that the slip length of the liquid
over the smooth hydrophobic solid is about 20~nm. Using the
macroscopic approach, we can represent the evolution of the
macroscopic slip length $B$ as a function of $\zeta$. Since $B$ is
required to be the largest, one can naively think that we should
have the largest possible $\zeta$. However, as we have seen with
the molecular dynamic simulations, high slippage requires the
liquid to be in a ``super-hydrophobic'' state, so that it can
``dewet'' over the roughness, gi\-ving infinite slip length at the
liquid-vapor interface. For this, the pressure has to be roughly
lower than the ca\-pil\-la\-ry pressure (as long as the grooves are deep
enough). This capillary pressure can be expressed in terms of
$\zeta$ : $P_{cap}=\frac{-2\gamma_{lv}\cos{\theta}}{\zeta L}$. If we consider
a pressure drop ex\-pe\-ri\-ment in a microfluidic channel, the pressure
drop that has to be applied is typically of $0.4\cdot
10^5$~Pa~\cite{cheng2002}. As re\-pre\-sen\-ted on
figure~\ref{fig:Pzeta} this gives us the upper limit of $\zeta$
($\zeta=0.88$ in our example). Therefore, if we consider grooves
of 100~nm width and periodicity 1~$\mu$m one may expect slip
length as large as 600~nm (figure~\ref{fig:Bzeta}). This
description does not determine the height of the grooves, but the
main constraint for them is to be high enough, so that the
capillary pressure does actually tune the transition between the
normal and the super-hydrophobic state (of course their height is
also limited by the technological fabrication possibilities);
typically here a height of 500~nm should be appropriate.

 In summary, use of surfaces patterned at the nanometer scale, and
 treated to produce a 'water repellent' like effect, appears to be
 a promising way towards devices that would allow flow of liquids
 in small size channels with very small flow resistance.

 \acknowledgments It is
a pleasure to thank C. Ybert for helpful conversations. We thank
the D.G.A. for its financial support. Numerical calculations were
carried out on PSMN (ENS-Lyon), CDCSP and P2CHPD (University of
Lyon) computers.

\section*{Appendix: calculation details}

Here we present in some detail the resolution of the stationary
Stokes equation with a locally imposed  slipping h.b.c.. We have
defined in equation~(\ref{defV}) the relative velocity  field :
$${\bf V}={\bf U}-\dot{\gamma}z\bf{e}_x$$
 which verifies  the following boundary conditions :\\
\\
$\bullet$ \underline{For $z=0$} :
\begin{equation}
\left\{
\begin{array}{rl}
{b(x,y)}\left(\left.\frac{\partial{V}_x}{\partial z}\right|_{z=0}+\dot{\gamma}\right) & ={V}_x(x;y;z=0)\\
\\
{b(x,y)}\left.\frac{\partial{V}_y}{\partial z}\right|_{z=0} &
={V}_y(x;y;z=0)\\
\\
{V}_z(x;y;z=0) & =0
\end{array}
 \right.
 \label{eqclz}
\end{equation}
\\

\noindent$\bullet$ \underline{For $z\ \rightarrow\infty$} :
\begin{equation}
{\bf V}=\alpha\bf{e}_x+\beta{\bf e}_y \label{eqclzinfty}
\end{equation}
\\

We solve this problem by expanding the various fields in Fourier
series, assuming for simplicity the dimensionless variables as
defined in (\ref{eq:adimension}) and a pattern with periodicity
2$\pi$ in both the $x$ and $y$ direction. For that we use the
following definition for the Fourier coefficients in $x$ and $y$
directions
\begin{equation}
\hat{{\bf V}}(q_x;q_y;z)=\frac{1}{(2\pi)^2 }\int\int_{[-\pi
;\pi]}e^{-i{\bf q}. \bf{r}}\bf{V}(x;y;z)dxdy
\end{equation}
where  ${\bf q}$ is a two dimensional wave vector, ${\bf
q}=q_x\hat{{\bf e}_x}+q_y\hat{{\bf e}_y}$ with discretized values
$q_{x,y}=n_{x,y}$. The vorticity ${\bomega}={\nabla}\times{\bf V}$
can then be written in Fourier space as :
\begin{equation}
\left\{
\begin{array}{rl}
\hat{\omega}_x & = iq_y\hat{V}_z-\frac{\partial \hat{V}_y}{\partial z}\\
\\
\hat{\omega}_y & = \frac{\partial \hat{V}_x}{\partial z}-iq_x\hat{V}_z\\
\\
\hat{\omega}_z & = iq_x\hat{V}_y-iq_y\hat{V}_x\\
\end{array}
 \right.
 \label{eq:syst1}
\end{equation}

We can deduce from equation~(\ref{eq:laplome}) and from the fact
that $\bomega$  must vanish as $z\rightarrow\infty$, that :
\begin{equation}
\left\{
\begin{array}{rl}
\hat{\omega}_x({\bf q};z) & = \underline{\hat{\omega}_x}({\bf q})e^{-qz} \\
\\
\hat{\omega}_y({\bf q};z) & = \underline{\hat{\omega}_y}({\bf q})e^{-qz} \\
\\
\hat{\omega}_z({\bf q};z) & = \underline{\hat{\omega}_z}({\bf q})e^{-qz} \\
\\
\end{array}
 \right.
 \label{eq:syst2}
\end{equation}

By taking the Fourier transform of the h.b.c.~(\ref{eqclz}), and
taking into account that $\hat{V}_z(z=0)=0$, we obtain the system
(\ref{eq:sysnis}) for $q\neq
0$.\\
\begin{equation}
\left\{
\begin{array}{rl}
\left[\hat{b}\circ\underline{\hat{\omega}}_y\right]({\bf
q})+\hat{b}({\bf q}) &=\frac{q_x}{2q^3}\left(q_x
\underline{\hat{\omega}}_y({\bf q})-q_y\underline{\hat{\omega}}_x({\bf q})\right)-\frac{\underline{\hat{\omega}}_y({\bf q})}{q}\\
-\left[\hat{b}\circ\underline{\hat{\omega}}_x\right]({\bf
q}) & =\frac{q_y}{2q^3}\left(q_x
\underline{\hat{\omega}}_y({\bf
q})-q_y\underline{\hat{\omega}}_x({\bf
q})\right)+\frac{\underline{\hat{\omega}}_x({\bf q})}{q}
\end{array}
\right. \label{eq:sysnis}
\end{equation}

The Fourier transform of (\ref{eqclz}) gives for $q=0$
\begin{equation}
\left\{
\begin{array}{rl}
\left[\hat{b}\circ\hat{\omega}_y\right]({\bf
q}=\bf{0})+\hat{b}({\bf q}=\bf{0})
& = \alpha \\
-\left[\hat{b} \circ \hat{\omega}_x\right]({\bf q}=\bf{0}) & = \beta \\
\end{array}
 \right.
 \label{eqsol}
\end{equation}
where $\circ$ denotes a convolution sum.

 By solving~(\ref{eq:sysnis}) for ${\omega}_x$ and
${\omega}_y$ and replacing into~(\ref{eqsol}), we can determine
$\alpha$ and $\beta$ and then the macroscopic slip length.
Equations~(\ref{eq:sysnis}) only involve
$\underline{\hat{\omega}}_x$, $\underline{\hat{\omega}}_y$ and
$\hat{b}$. Since $\hat{b}$ is an input of the problem, it is then
possible to determine $\underline{\hat{\omega}}_x$ and
$\underline{\hat{\omega}}_y$.

This resolution of~$\underline{\hat{\omega}}_x$ and
$\underline{\hat{\omega}}_y$ can not usually be done analytically
 (except for very simple patterns of $b$) and a numerical approach is required.
For that, we consider a discretization of the elementary cell (of
size $(2\pi)^2$ in the real space) into $N^2$ nodes of coordinates
:
\begin{equation}
\begin{split}
(x_k;y_k)=\left(-\pi+k_x\frac{2\pi}{N};-\pi+k_y\frac{2\pi}{N}\right)\\
\quad\text{where}\quad(k_x,k_y) \in[0,N-1]^2
\end{split}
\end{equation}
The discretisation of real space implies periodicity in the
Fourier space with period $N$. We then only consider the wave
vectors ${\bf q}$ in the interval :
$$
(q_{x};q_{y})\in[0,(N-1)]^2
$$
By introducing the complex vector $\hat{\Omega}({\bf
q})=\underline{\hat{\omega}}_x({\bf
q})+j\underline{\hat{\omega}}_y({\bf q})$ and by making a
judicious combination of the equations~(\ref{eq:sysnis}), the
problem is reduced to a matrix system of the form :
\begin{equation}
\left(
\begin{array}{c}
M
\end{array}
\right) \left(
\begin{array}{c}
\Omega
\end{array}
\right)= \left(
\begin{array}{c}
B
\end{array}
\right) \label{eqmat}
\end{equation}
where $\left(\Omega\right)$ is a vector of size 2($N^2$-1)
involving only $\underline{\hat{\omega}}_x$ and
$\underline{\hat{\omega}}_y$ , and $\left(B\right)$ is a vector of
size 2($N^2$-1) that only involves the prescribed microscopic slip
length $b$. The size of the matrix $\left(M\right)$ is
4$\times$($N^2$-1)$^2$, which increases rapidly with $N$. However,
for most of the situation of interest, ($M$) remains a sparse
matrix with many zero coefficients. The density of the matrix does
actually depend on the pattern of the microscopic h.b.c.. For a
strip pattern of microscopic h.b.c., for example, only 3~\% of the
elements of the matrix' elements are non zero (this spareness
actually results from the invariance by translation of the pattern
in the direction of the stripes). This allows the implementation
of efficient sparse matrix algorithms to solve
equation~(\ref{eqmat}). We typically consider N up to N=256 grid
points both in $x$ and $y$ directions. The matrix (M) is then
inverted using a conjugated gradient method well suited to the inversion of matrix of low densities.\\

\begin{center}

\begin{table}
\begin{center}
\begin{tabular}{|c|c|}
\hline ``Real'' value of $\zeta$  & Value of $\zeta$ from the\\
&linear fit of $B$ ($\pm~2\%
$)\\
&as a function of $b_1$\\ \hline 97&96.7\\
\hline
83.3& 83.2\\
\hline
50 & 49.2\\
\hline
25 &24.8\\
\hline
10 &9.9\\
\hline

\end{tabular}
\end{center}
\caption{Comparison of the ``real'' value of the fraction $\zeta$
and the value one get with a linear fit using the
equation~\ref{eqBb}} \label{tablezeta}
\end{table}
\end{center}

\end{document}